# Strain-stiffening gels based on latent crosslinking

Yen H. Tran[1], Matthew J. Rasmuson[1], Todd S. Emrick[2], John Klier[1*], Shelly R. Peyton[1*]

[1]Department of Chemical Engineering
[2]Department of Polymer Science and Engineering

University of Massachusetts-Amherst, Amherst, MA 01003-9303, USA.

*Correspondence to:
John Klier
Department of Chemical Engineering
University of Massachusetts Amherst
159C Goessmann Laboratory
686 North Pleasant Street
Amherst, MA 01003-9303
Phone: (413)-545-2819
Email: klier@umass.edu

Shelly R. Peyton
Department of Chemical Engineering
University of Massachusetts Amherst
Life Sciences Laboratory
240 Thatcher Way
Amherst, MA 01003-9303
Phone: (413)-545-1133
Email: speyton@ecs.umass.edu

**Abstract**

Gels are an increasingly important class of soft materials with applications ranging from regenerative medicine to commodity materials. A major drawback of gels is their relative mechanical weakness, which worsens further under strain. We report a new class of responsive gels with latent crosslinking moieties that exhibit strain-stiffening behavior. This property results from the lability of disulfides, initially isolated in a protected state, then activated to crosslink on-demand. The active thiol groups are induced to form inter-chain crosslinks when subjected to mechanical compression, resulting in a gel that strengthens under strain. Molecular shielding design elements regulate the strain-sensitivity and spontaneous crosslinking tendencies of the polymer network. These strain-responsive gels represent a rational design of new advanced materials with on-demand stiffening properties with potential applications in elastomers, adhesives, foams, films, and fibers.

**Introduction**

Polymer gels are swellable, insoluble networks used in a broad array of applications, from tissue engineering to consumer products[1]. A key feature of gels is their mechanical tunability, wherein the elastic moduli can range from tens of Pa to tens of MPa[2]. The moduli of gels are controllable through alteration of backbone composition and crosslinking density[3-5]. Most important for biological applications, gels can be configured to encompass the soft and wet nature of living tissue[6]. Two major limitations in this field are that, once synthesized, the modulus of a gel network is fixed, and gels typically weaken under strain. This latter feature limits their utility in industrial applications that require robust strength during processing.

In contrast, nature's polymers exhibit strain-stiffening mechanisms[7-10]. Fibrin during blood clotting[11,12], and actin cytoskeletal filaments during cell movement[13], stiffen rapidly under deformation[14]. These biological polymers inspire the creation of strain-responsive synthetic materials. Strain-induced strengthening has been widely exploited to improve the performance of several traditional polymer classes. For example, thermoplastics for application as films, fibers, and foams are tunable via force-induced molecular orientation and crystallization/physical crosslinking[15,16]. Realizing similar mechanical tuning of gels would greatly broaden their potential and utility.

Recent innovations have created dynamically responsive gels with increasing moduli induced via heat or light[17-23], but these typically require sophisticated chemistries, and would not work for materials in dark environments, such as adhesives. Attempts to selectively tailor the strength and functionality of synthetic gel polymer networks by exploiting mechanical deformation have resulted in transient gel stiffening due to chain associations or elasticity, but haven not achieved permanent strain-induced modulus or crosslinking increases across the range of deformations[23-30]. Subsequent gel deformation in these systems typically produced similar or lower stiffness than the initial deformation, suggesting that the bonds are meta-stable, in contrast to physical or chemical crosslinks.

While gels with strain-stiffening domains have not been described previously, related synthetic work provides a conceptual basis for such systems[31,32]. We describe novel gels containing latent crosslinking domains based on a thiol-containing monomer within a poly(ethylene glycol) (PEG)-acrylate backbone. To our knowledge, this is the first report of gels that employ latent crosslinking to promote on-demand, strain-induced hardening while using simple, scalable polymer chemistry. For many gel applications, strain-induced crosslinking activation could provide a new type of latency, delaying crosslinking until the onset of mechanical activation[33-35].

**Design of strain-responsive gels**

Inspired by the force-induced arrangement of disulfide bonds in proteins[36,37], we developed a self-reinforcing material reliant on reversible chemical crosslinks. A dihydrolipoic acid (DHLA)-based methacrylate was integrated into the networks to impart reversible inter-chain disulfide crosslinks[31,32]. As a typical example, gels were prepared from an acrylate polymer containing PEG and lipoic acid (LA) groups (i.e., from PEG acrylate and 2-hydroxyethyl methacrylate-lipoic acid, HEMA-LA, Fig. 1a and Supplementary Fig. S1). The disulfides of the Las serve as latent thiol sources, activated upon chemical reduction and manifested upon applications of force. Such materials represent synthetic variants of 'cryptic sites' in proteins, in which reactive or binding groups are initially shielded sterically (i.e., in folded regions) but are revealed upon force-induced unfolding or protein degradation[38,39]. The effects of gel structure and composition were investigated by varying crosslinker and co-monomer chain lengths across material Sets I-III (Fig. 1b) and, within each set, by varying monomer composition across Groups A-C (Fig. 1c). We denote the samples using the code X-Y, where X is the gel set and Y is the gel group. At a fixed crosslinker concentration of 0.129 M, these gels were transparent with representative gels from Set I shown in Fig. 1d. Schematic illustrations of molecular changes in gel network during disulfide reduction, followed by mechanical deformation, are shown as a function of co-monomer (M1) molecular weight (or chain length) (Fig. 1e).

To prepare the gels, we first synthesized networks of X-Y composition. The cyclic disulfides were reduced to pendant thiols by adding a dilute solution of sodium borohydride to the gels. This would theoretically facilitate inter-chain disulfide reactions to form new inter-chain crosslinks after reaction of pendant thiol group, a process which may be accelerated by the application of stress to the network (Fig. 1e). Moreover, variation of M1 chain length could influence the strain-stiffening behavior of reduced gels, in which short M1 groups would more easily allow inter-chain crosslinking under force while long M1 groups might shield free thiols from crosslinking by disulfide formation.

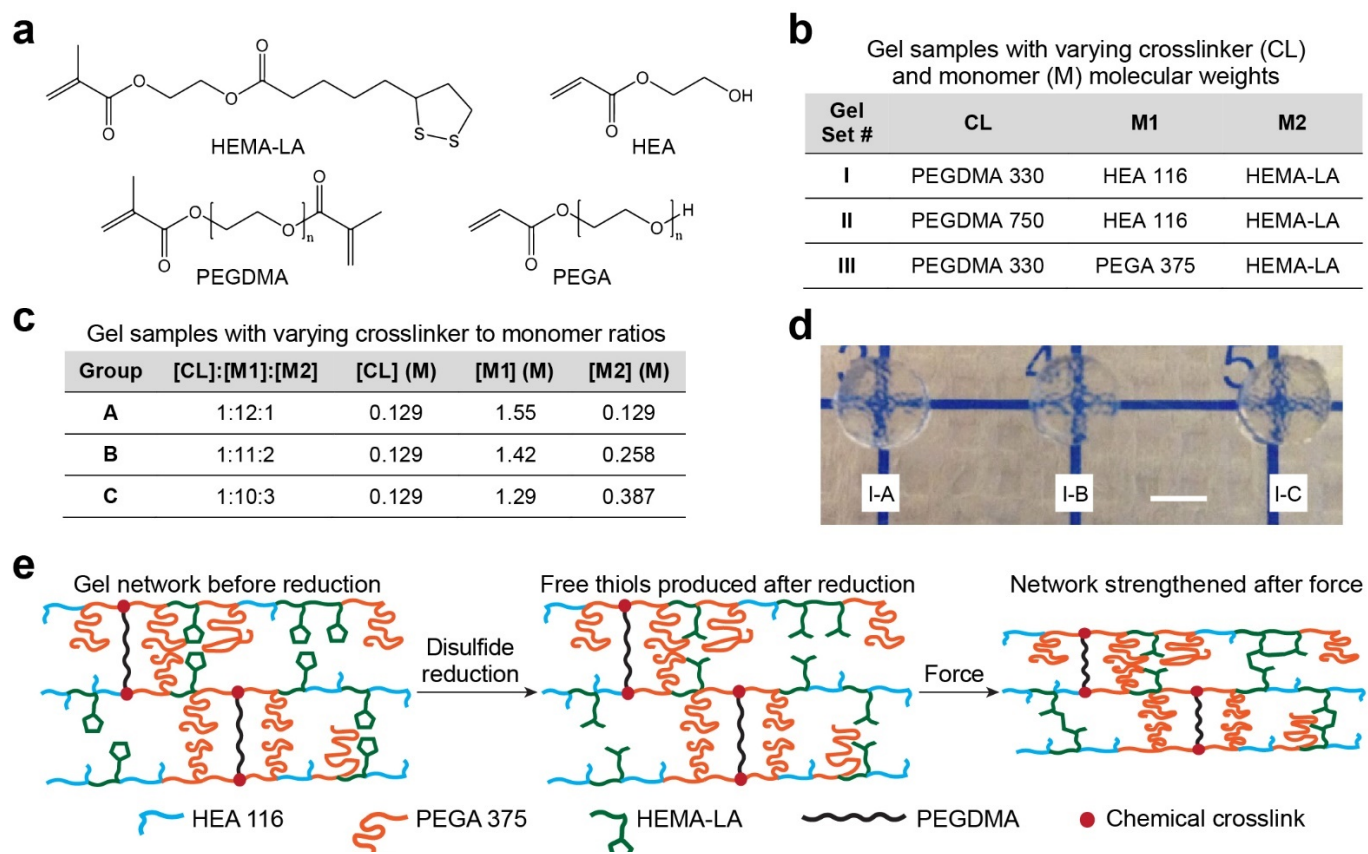

**Figure 1 | Design of strain-responsive gels. a,** Monomer and crosslinker structures. **b,** Table detailing three sets of gels (I-III), varying the chain lengths of crosslinker (CL) and co-monomer (M1). **c,** For each set of gels, three groups with different compositions were used, where crosslinker concentration and total monomer concentration [M1]+[M2] were kept constant. **d**, Photographs of optically clear gels from Set I after swelling for 5 days in 1:4 (v:v) ethanol:DMSO. Scale bar: 3 mm. **e,** Illustration of gel networks of different co-monomer molecular weights. Under applied force, short M1 groups (blue) allow inter-chain crosslinking while long M1 groups (orange) limit disulfide formation.

**Spontaneous crosslinking**

After disulfide reduction, gels were immersed in 1:4 (v:v) ethanol : dimethyl sulfoxide (DMSO) under ambient conditions to promote crosslinking by oxidation of thiols to disulfides. We characterized gel performance such as optimization of reduction time and yield, thiol quantification, and swelling behavior (Supplementary Fig. S2). The optimal reduction time was determined to be 6 hours. Gels deswelled about 1 day after the reduction, indicating that thiol crosslinking occurred over this timeframe. Free thiol concentration was quantified by Ellman's assay[40], and Young's modulus was measured by compression testing at various time points post-reduction. Within 1 hour, the changes in thiol concentration were negligible in all groups of gels from Set II (Fig. 2a), and no changes in stiffness were observed (Fig. 2b). Within 5 days, significant declines in thiol concentration due to disulfide formation accompanied increased gel stiffness (Fig. 2c), indicating that some proportion of the disulfide bonds provided inter-chain crosslinks (Fig. 2d and Fig. 1e).

An increase in gel stiffness correlated to a decrease in free thiol concentration associated with disulfide formation and crosslinking. Since Flory has described the relationship between elastic modulus and degree of crosslinking[41], we compared Flory's model (Equation 1) to our data on available thiols to measured modulus (Fig. 2e).

$$G = \frac{\rho RT}{M_c}\left(1 - \frac{2M_c}{M}\right) \qquad (1)$$

where $G$ is the shear modulus, $\rho$ is the density of the network, $R$ is the gas constant per mole, $T$ is the temperature, $M_c$ is the mean molecular weight of the chains, and $M$ is the molecular weight of the primary molecules before crosslinking. Given that our data yielded less than the ideal 100% conversion assumed by Flory, we expect spontaneous crosslinking resulted in both inter- and intra-chain disulfide bonds. Group II-C showed lower conversion of inter-chain disulfides compared to Groups II-A and II-B, possibly due to its higher initial thiol density, creating thiol crowding and reformation of cyclic disulfides.

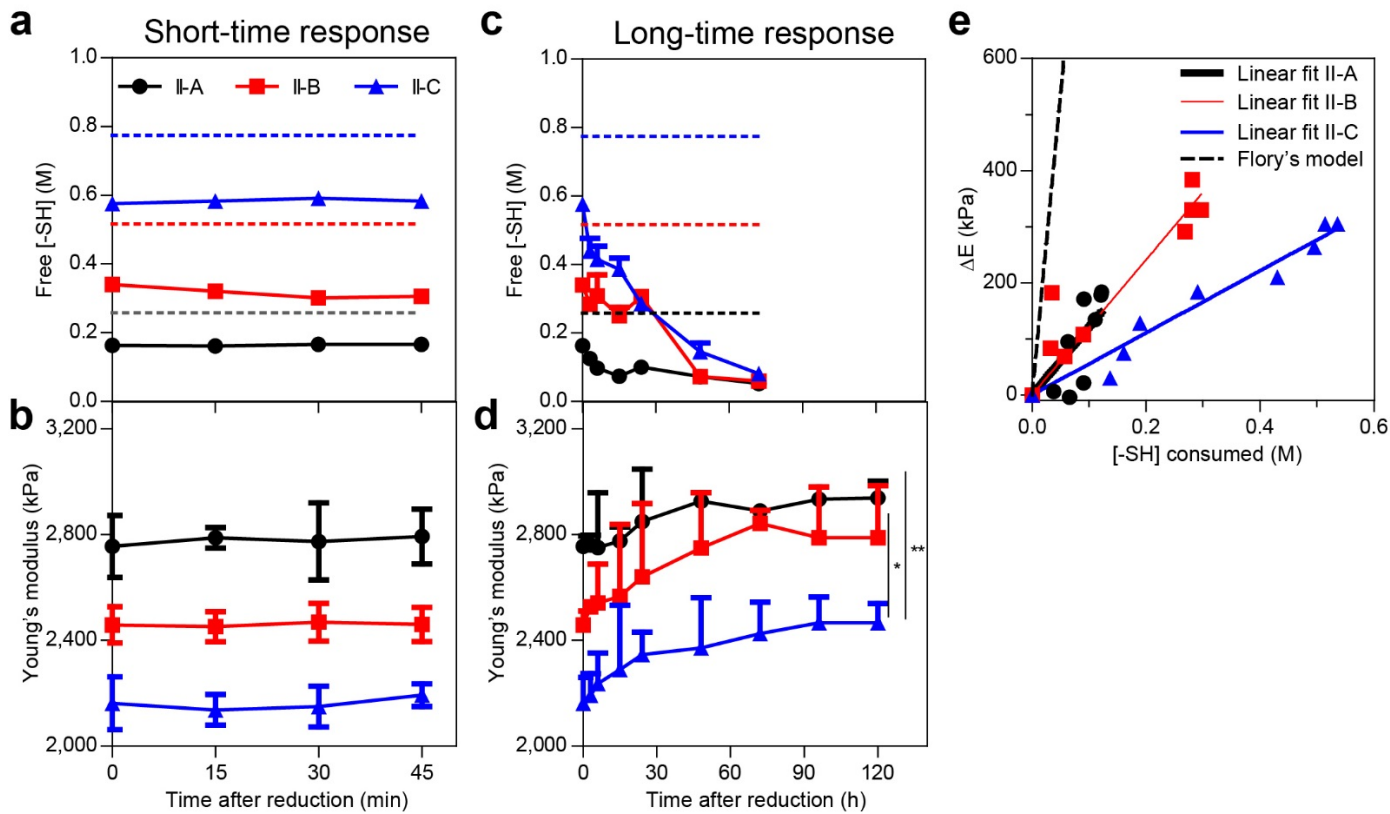

**Figure 2 | Strain-induced stiffening network through labile thiol crosslinks. a-d,** Free thiol concentration (top) and Young's modulus (bottom) as functions of time post-reduction. Solid lines: experimental data; dashed lines: theoretical thiol concentration at 100% yield of reduction. **a-b,** Short-time response (45 min). **c-d,** Long-time response (5 days). **e,** Correlation between the increase in stiffness of reduced gels and thiol consumption. Solid lines: experimental data with their corresponding linear fits; dashed line: Flory's theory of elastic model. Note that the linear fits for II-A (black) and II-B (red) are superimposed, thus the linear fit for II-A is made thicker for visualization purposes.

**Reversibility of latent thiol crosslinking**

Due to the labile nature of disulfide crosslinks, we expected that they would be reversible and amenable to repeated stiffening and weakening by toggling the oxidation-reduction reactions[42]. Indeed, the gels showed complete reversibility of the disulfide crosslinks and Young's modulus over three cycles (Fig. 3). We first reduced gels to convert the cyclic disulfides to free thiols (Day 1). Thiol-containing gels were then allowed to form intra-chain or inter-chain disulfides through oxidation by continuous purging of air into the system. Gel stiffness and thiol concentration were measured daily for 4 days until most thiols were oxidized, and no significant change in stiffness was observed. The second (Day 5) and third (Day 9) cycles were carried out in the same manner on the same gels. Gels underwent multiple reduction-oxidation cycles without deterioration, highlighting the reversibility of the process. In fact, thiol concentration and Young's modulus of gels after each reduction showed effectively 100% recovery, i.e. no significant differences in thiol concentration or stiffness were observed on Days 1, 5, and 9. This reversible behavior is distinct from previous reports of covalently crosslinked gels, which showed complete crosslinking/decrosslinking of all covalent bonds in the network via sol-gel transition[43-45].

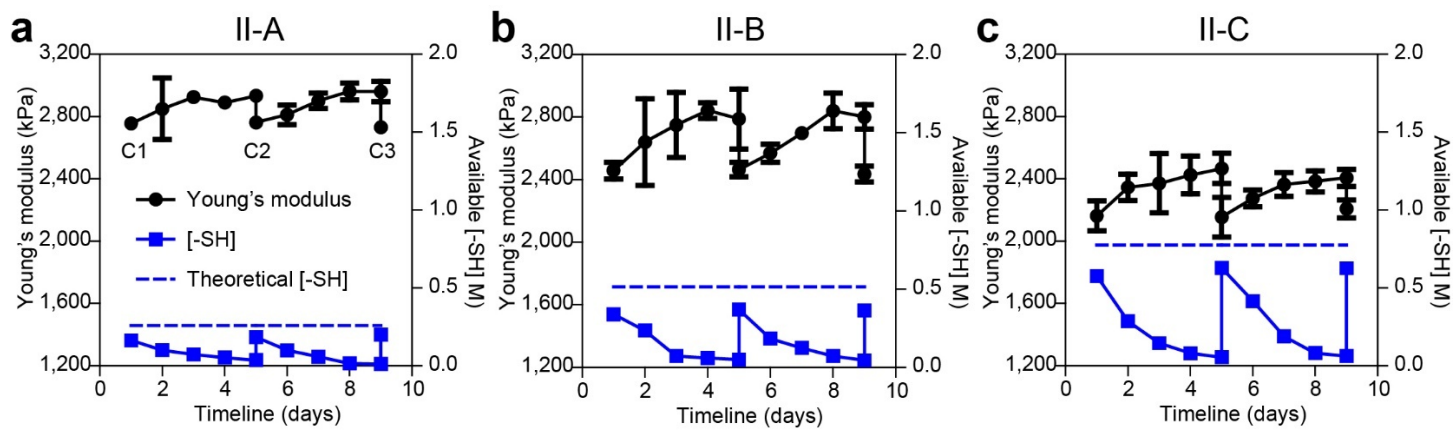

**Figure 3 | Reversibility of latent thiol crosslinking. a-c,** Recovery of thiols (blue) and Young's modulus (black) over three oxidation-reduction exchange reactions from Set II with different gel compositions. Cycles 1, 2, and 3 started on Days 1, 5, and 9, respectively. **a,** II-A (low [-SH]). **b,** II-B (intermediate [-SH]). **c,** II-C (high [-SH]).

**Molecular shielding of gel stiffening**

Crosslinker and co-monomer compositions were manipulated to control spontaneous crosslinking behavior. The M1 molecular weights were 116 and 375 g/mol, while the crosslinker molecular weights were 330 and 750 g/mol (Fig. 1b). As shown in Fig. 4, gels increased in stiffness at a modest rate over time for the combinations of short CL/short M1 used in Set I (Fig. 4a) and of long CL/short M1 used in Set II (Fig. 4b), but a smaller increase in stiffness was observed in the combination of short CL/long M1 used in Set III (Fig. 4c). These results indicated that gels composed of long co-monomer M1 provided "screening" of spontaneous inter-chain crosslinking while short M1 allowed more rapid intermolecular disulfide bond formation (Fig. 1e).

Crosslinker chain length did not significantly affect the formation of new, spontaneous crosslinks. The molar ratio of crosslinker to total monomer integrated in all groups of gels was 1:13, meaning the crosslinker had little control over thiol activity. Based on literature reports, we expected the disulfide bonds to remain stable under tensions of several hundred pN[36,37], which is larger than those based on hydrophobicity, but still well below the forces of several nN required to rupture C-C bonds in the polymer backbone[46,47]. After the reduction, the expectation was that some pendant thiols would form inter-chain crosslinks leading to stiffer gels. Hence, stiffness could be increased further by placing the gels under compression, accelerating inter-chain thiol crosslinking instead of reforming the cyclic disulfides. In contrast, long M1 slowed inter-chain crosslinking, and gave gels with no detectable changes in stiffness (Fig. 1e). Instead, the presence of these shielding groups biased the system toward cyclic disulfide reformation and no thiol was detected in the oxidized gels.

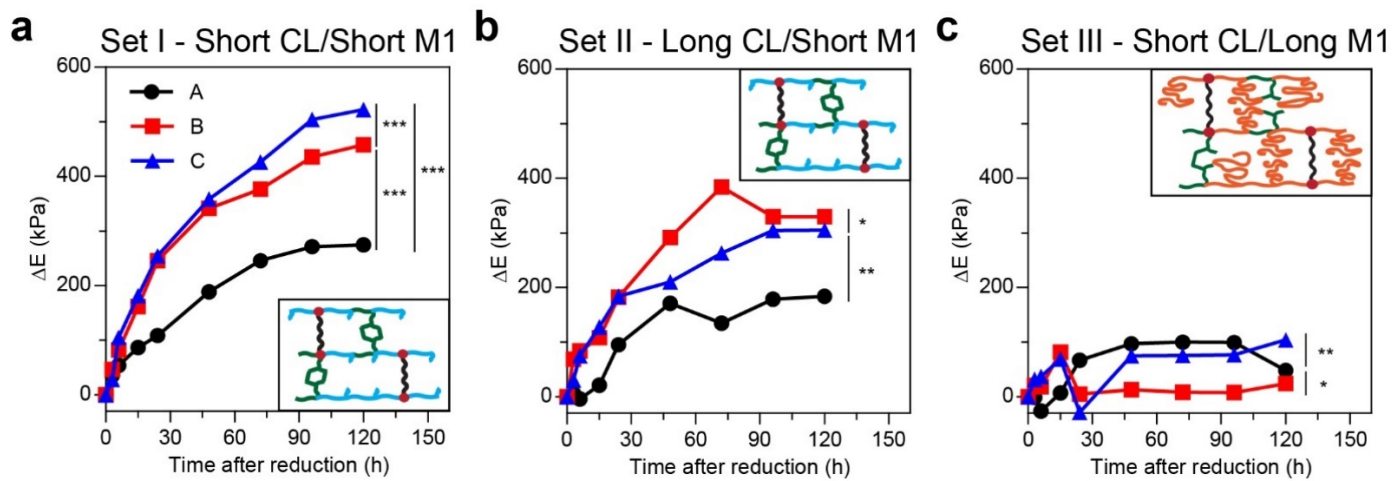

**Figure 4 | Shielding effect on gel stiffening via long dangling monomers. a-c,** Change in Young's modulus ΔE over time post-reduction of different gel structures. **a,** Set I (short CL/short M1). **b,** Set II (long CL/short M1). **c,** Set III (short CL/long M1). Insets: schematics of disulfide formation in gel network during oxidation.

**External strain accelerates inter-chain crosslinking**

Besides spontaneous crosslinking, labile thiols in the networks were also responsive to mechanical strain. The increase in Young's modulus of strain-stiffening gels accelerated with applied stress. Under compression, inter-chain crosslinking resulted in faster stiffening than was observed in the spontaneous state (Fig. 5). At equivalent times post-reduction, the increase in Young's moduli under strain was greater than those of unstrained gels. We hypothesize that under deformation, the reactive groups were brought into close proximity, accelerating inter-chain crosslinking and stiffening. Gels with lower thiol content (Fig. 5a, b) exhibited smaller increases in stiffness (Fig. 5c), since fewer crosslinks produce weaker networks. This result indicated that the strain-responsive nature of the gels allows accurate control over stiffness through thiol density.

These experimental results were compared to Flory's model of elastic systems using gels from Set II (composed of long CL and short M1). For all groups in Set II, the increase in stiffness from gels under strain approached Flory's model more closely than unstrained gels. Reduced gels with higher initial free thiol concentration gave greater increases in Young's modulus, ranging from 150 kPa in a low thiol concentration (II-A) to 600 kPa in a high thiol concentration (II-C). Moreover, most inter-chain thiol crosslinking occurred during the first 2 hours, then slowed when most of the available thiols were consumed (Fig. 5). Notably, the increase in stiffness of II-C gels under strain after 3 hours was greater than that of gels without strain after 5 days. As noted in Fig. 2e, II-C gels of highest thiol concentration tended to reform cyclic disulfides, resulting in the smallest increase in stiffness at a given thiol consumption. However, under cyclic deformation, these free thiols formed inter-chain disulfides faster than the unstrained gels.

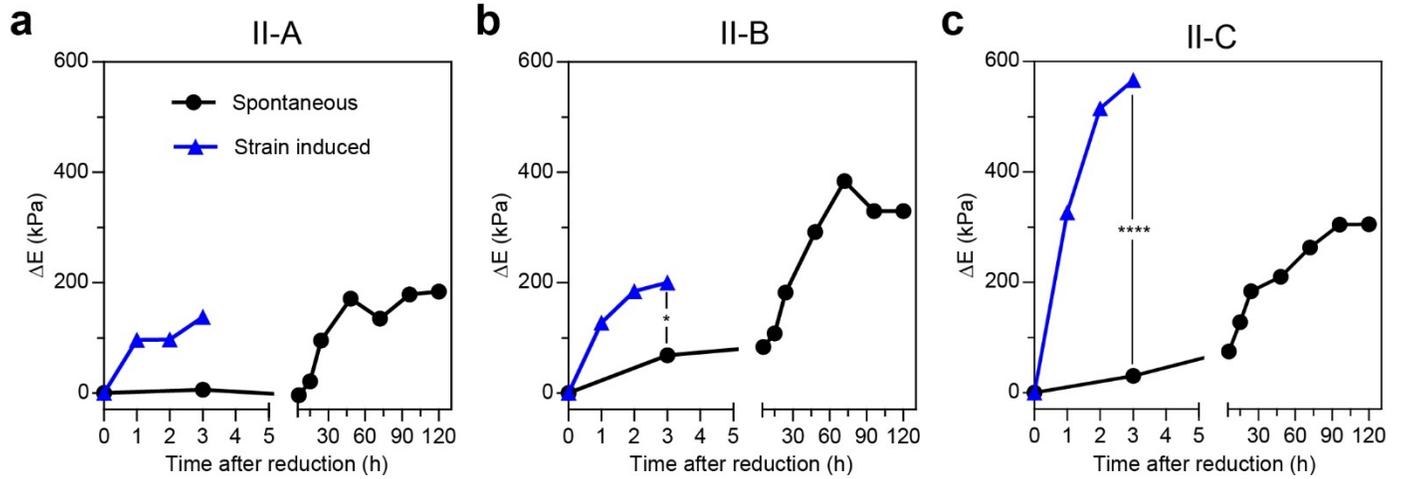

**Figure 5 | Strain-induced latency of intermolecular disulfide crosslinks. a-c,** Change in Young's modulus post-reduction under strain (blue) for 3 hours compared to spontaneous state (black) using Set II. **a,** II-A (low [-SH]). **b,** II-B (intermediate [-SH]). **c,** II-C (high [-SH]).

**Suppression of strain-stiffening via molecular shielding**

We further investigated gel stiffening by covariation between two factors, co-monomer M1 chain length and crosslinker concentration; each was varied between two levels, denoted as - for low levels and + for high levels (Fig. 6a). We tested four gel compositions, denoted as L1/L2, where L1 and L2 were the levels for the M1 molecular weight and crosslinker concentration, respectively (Fig. 6a). The levels for M1 molecular weight were 116 g/mol (HEA) and 950 g/mol (PEG methyl ether methacrylate, PEGMEMA), and those for crosslinker concentration were 0.065 M and 0.129 M. The crosslinker was PEG diacrylate (PEGDA, average $M_n$ 700), and the concentrations of M1 and M2 (HEMA-LA) were fixed at 1.29 M and 0.387 M, respectively. For groups that contained the same crosslinker concentration, i.e. +/- and -/-, or +/+ and -/+, the Young's moduli of gels immediately after reduction (t=0), termed "starting moduli," were similar due to equal volume fraction of polymer in the gels (Fig. 6b). However, the stiffening behavior over time was only observed in gels with shorter M1, i.e. -/- and -/+, whereas no significant increase in stiffness occurred in gels with longer M1, i.e. +/+ and +/-. This behavior was evident by the change in stiffness as a function of time post-reduction (Fig. 6c). As depicted in Fig. 1e, the short co-monomer M1 allowed for strain-induced formation of new crosslinks between inter-chain thiols, while the long M1 groups acted as molecular shielding units, inhibiting inter-chain crosslinks even under applied force.

To determine whether the shielding effect could be overcome under strain, we selected the +/- gel to perform a cyclic compression test over 18 hours (Fig. 6d). The shielding effect resulting from bulky co-monomer M1 persisted under strain during the first 3 hours. However, as stress was applied, the gel gradually stiffened by intermolecular crosslinking, giving this strain-induced gel a similarly large increase in Young's modulus to the gels composed of short M1 (Sets I and II) at a spontaneous state. This result could be used to develop a cryptic network that minimizes spontaneous stiffening but is accessible and stiffened during strain[48-50].

In summary, we have demonstrated a strain-triggered stiffening synthetic gel network. We achieved this with polymers containing labile thiol crosslinks on an acrylate backbone with pendant PEG chains, which allowed for both spontaneous and strain-sensitive crosslinking. The labile nature of these thiol groups also allowed for reversibility of the crosslinks. Future studies will include applications where crosslinking is induced via mechanical stimulation, such as shaking, compression, or ultrasound.

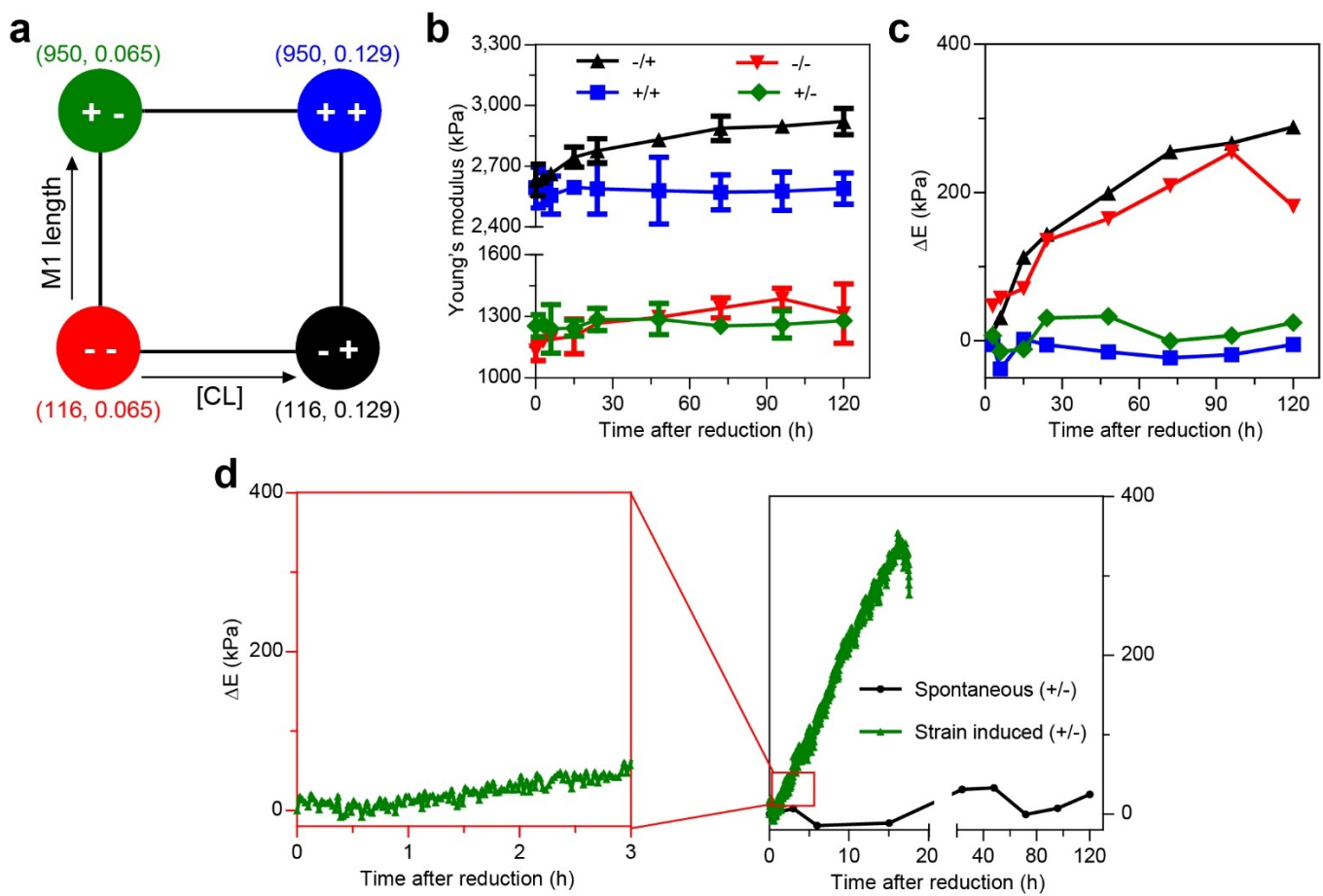

**Figure 6 | Suppression of shielding effect from long monomers under long-term deformation. a,** Illustration of experimental design using co-variation of two independent variables, co-monomer chain length and crosslinker concentration, each varied between two levels called low and high levels, denoted as - and +, respectively. **b,** Time dependence of Young's modulus for different gel compositions listed in **a**. **c,** Change in Young's modulus post-reduction compared to the initial stiffness. **d,** Long-term strain stiffening on (+/-) gel over 3 hours (left) and 18 hours (right, green) compared to spontaneous state (black).

## Methods

Strain-stiffening gels were synthesized by random copolymerization of HEMA-LA and an acrylate-based co-monomer. A precursor solution with the prescribed compositions of monomers, crosslinker, and 0.8 wt% Irgacure 2959 free radical initiator in DMSO was degassed for 40 s under $N_2$ (g), poured into cylindrical Teflon molds (5x5 mm), and cured under 365-nm ultraviolet light for 30 min. After polymerization, the resulting gels were immersed in a large amount of 1:4 (v:v) ethanol:DMSO for 5 days, which was changed twice daily to remove any uncrosslinked materials and to allow the gels to equilibrate. For disulfide reduction, the gels were placed in a 12-well plate (one gel per well). The reducing solution was prepared by dissolving sodium borohydride (8 molar equiv relative to lipoic acid) in 1:4 (v:v) ethanol:DMSO solution. A volume of 3 mL of reducing solution was added to each well and shaken on an orbital shaker for 6 hours.

The available thiol concentration in the reduced gels was measured using Ellman's test. Post-swelling, compression stress-strain mechanical measurements were performed using a parallel plate rheometer at a strain rate of 20 um/s in air to characterize the elastic moduli of the gels. For long-term strain induction test, gels were immersed in a 35-mm Petri dish containing the swelling solution to avoid drying. The Young's modulus $E_m$ for each gel was calculated by plotting the measured normal force between 0-5% strain and dividing the slope of the best-fit linear regression by the gel cross-sectional area. To normalize mechanical stress measurement among the samples with different degrees of swelling, the stress applied to the solid matter in the gels $E_r$ as follows:

$$E_r = \frac{E_m}{r^{2/3}} \qquad (2)$$

Additional detailed methods are given in supplementary materials.

**Acknowledgements**

This research was financially supported by a CAREER award from the National Science Foundation (NSF) grant (DMR-153514) to S.R.P. and the University of Massachusetts (UMass) Amherst to J.K. S.R.P. is a Pew Biomedical Scholar supported by the Pew Charitable Trusts and a Barry and Afsaneh Siadat faculty award. We are thankful to the NSF Materials Research Science and Engineering Center at UMass (NSF DMR-0820506) for use of the core rheometer and NMR spectroscopy. We thank Dr. Pornpen Sae-ung for assistance with HEMA-LA synthesis and Dr. Jungwoo Lee for the use of the plate reader. We also thank Carey Dougan, Dr. Maria Gencoglu, Lauren Jansen, and Alyssa Schwartz for their thoughtful comments on the paper.

**Author contributions**

J.K. and S.R.P. conceived the project. T.S.E. conceived the synthesis of HEMA-LA monomer. Y.H.T., J.K. and S.R.P. designed the experiments. Y.H.T. and M.J.R. performed the experiments. Y.H.T. analysed the data. All authors contributed to writing, reviewing, and editing the manuscript.

**Competing financial interests**

The authors declare no competing financial interests.

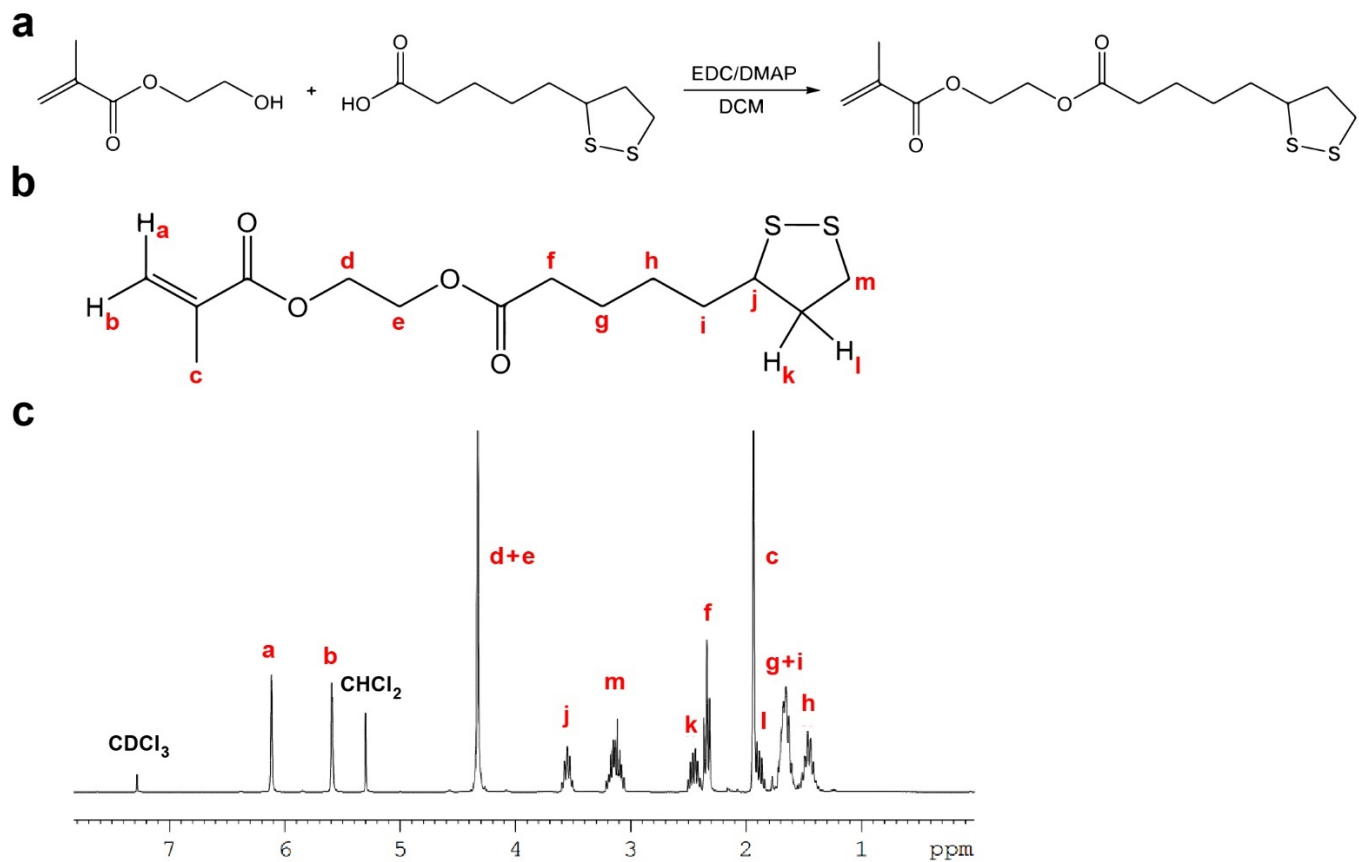

**Supplementary Figure 1 | Synthesis of HEMA-LA.** a, Homopolymerization of HEMA-LA by RAFT reaction between HEMA and LA using EDC as coupling agent and DMAP as catalyst in DCM. b, Chemical structure of HEMA-LA. c, $^{1}$H NMR spectrum of HEMA-LA.

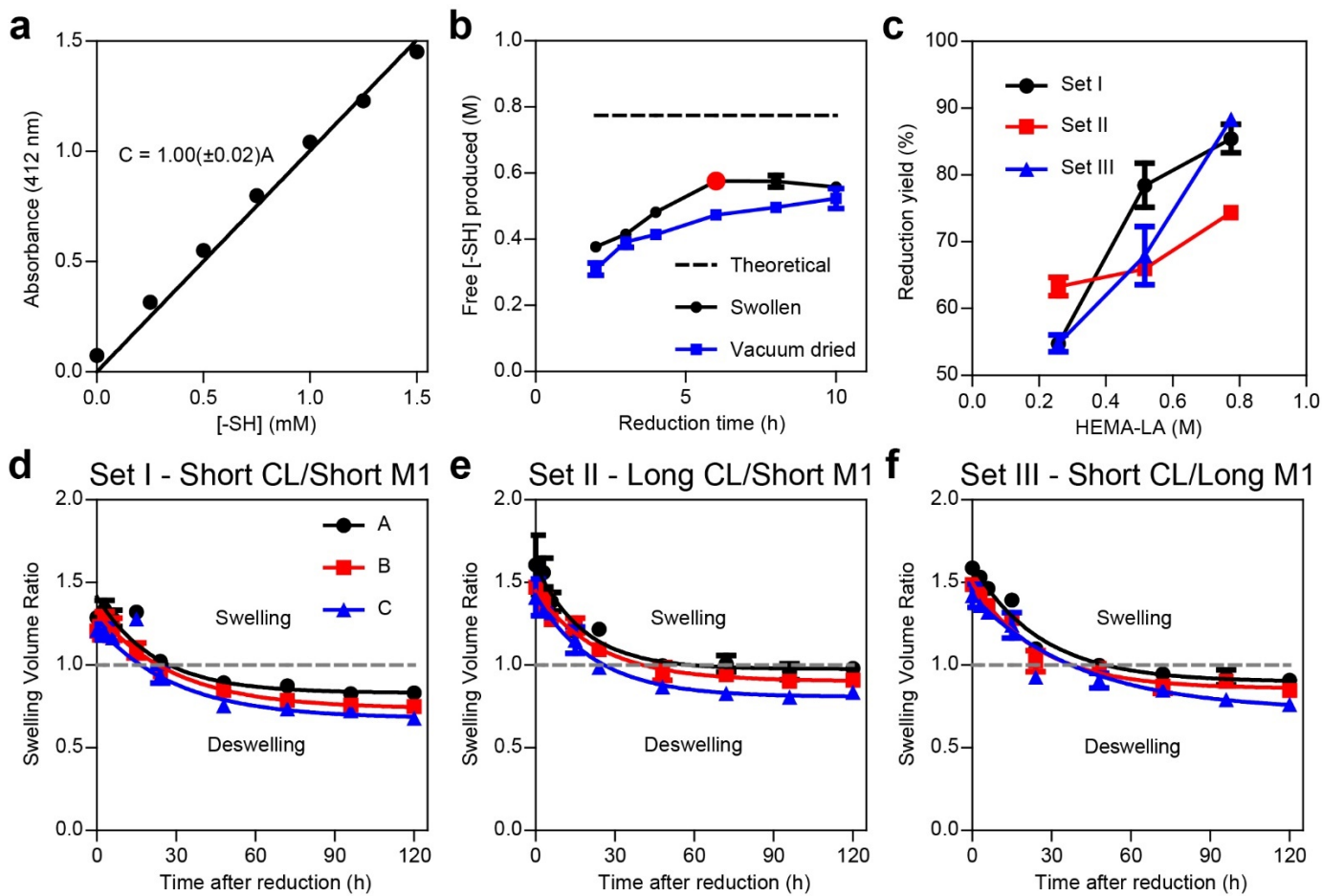

**Supplementary Figure 2 | Characterization of strain-stiffening gels. a**, Calibration of Ellman's reagent using cysteine standard to quantify thiol concentration. **b**, Optimization of reduction time using II-C gels with two different methods. **c**, Reaction yield after 6-hour disulfide reduction for different gel sets and gel groups. **d-e**, De-swelling behavior of HEMA-LA gels in 1:4 ethanol:DMSO as a function of time after reduction of different gel structures. **d,** Sets I (short CL/short M1). **e,** Set II (long CL/short M1). **f,** Set III (short CL/long M1).

*Methods*

**Synthesis of HEMA-LA.** HEMA-LA monomer was prepared following the previously reported method[1]. Briefly, 2-hydroxyethyl methacrylate (HEMA) (4.70 mL, 38.8 mmol), lipoic acid (LA) (8.00 g, 38.8 mmol), N-(3-dimethylaminopropyl)-N′-ethylcarbodiimide hydrochloride (EDC) (11.15 g, 58.2 mmol), and 4-(dimethylamino)pyridine (DMAP) (4.74 g, 38.8 mmol) were dissolved in 100 mL distilled dichloromethane in a dry round-bottom flask. The reaction mixture was stirred at room temperature for 24 h under nitrogen. The reaction was sequentially washed with 1 M $HCl_{(aq)}$, saturated $NaHCO_{3(aq)}$, and brine solutions. The organic layers were collected and dried over anhydrous $MgSO_4$, filtered, and concentrated by rotary evaporation, giving HEMA-LA monomer as a yellow oil (8.5 g, 70% yield). $^1$H NMR spectrum was recorded at 400 MHz on Avance400 spectrometer. $^1$H NMR (400 MHz, $CDCl_3$): δ = 6.06 (s, 1H), 5.36 (s, 1H), 4.26 (s, 4H), 3.5 (m, 1 H), 3.11 (m, 2H), 2.40 (m, 1H), 2.3 (t, 2H), 1.87 (s, 3H), 1.35−1.70 (m, 8H).

*Characterization of gels*

**Swelling studies.** The swelling volume ratio $Q$ is defined as the ratio of the sample volume at swelling equilibrium $V$ to the initial volume $V_0$. Swelling equilibrium is defined after gels sit in a swelling solution for 5 days. The weight fraction of polymer (polymer content) of swollen gels $r$ was determined as the ratio of the dry weight of the sample $W_d$ to its swollen weight $W_s$:

$$r = \frac{W_d}{W_s} \quad (1)$$

**Thiol concentration measurement.** The available thiol concentration in reduced gels was monitored using adapted Ellman's assay[20]. Briefly, a reaction buffer containing 0.1 M sodium phosphate and 1 mM EDTA, pH 8.0 was prepared. An Ellman's reagent solution was prepared by dissolving 4 mg of Ellman's reagent in 1 mL of reaction buffer. Cysteine hydrochloride

monohydrate was used as the standard. A set of seven cysteine standards was prepared as follows:

**Supplementary Table 1: Cysteine standards for Ellman's assay**

| Standard | Volume of reaction buffer (mL) | Amount of cysteine | Final concentration (mM) |
|---|---|---|---|
| 1 | 5.00 | 1.317 mg | 1.50 |
| 2 | 0.2 | 1 mL of Standard 1 | 1.25 |
| 3 | 0.4 | 0.8 mL of Standard 1 | 1.00 |
| 4 | 0.6 | 0.6 mL of Standard 1 | 0.75 |
| 5 | 0.8 | 0.4 mL of Standard 1 | 0.50 |
| 6 | 1.0 | 0.2 mL of Standard 1 | 0.25 |
| 7 | 1.2 | 0 | 0.0 |

A small portion of gel after being crushed was used as sample whose thiol concentration was ensured to be in the working range of the standard curve (0.1-1.0 mM). A volume of 250 µL of each standard was added to separate test tubes. For gel samples, an additional volume of reaction buffer was added to each gel portion so that the total volume was 250 µL. In each test tube of standards or gel samples, 50 µL of Ellman's reagent solution and 2.5 mL of reaction buffer were added. The mixture was mixed and incubated at room temperature for 15 min, followed by absorbance measurement at 412 nm using BioTek Synergy H1 microplate reader.

*Statistical analysis*

Statistical analysis was performed using Prism v5.04 (GraphPad Software). Data are reported as mean ± standard deviation, unless otherwise noted. Statistical significance of the difference between pairs of means was evaluated by computing *P*-values with unpaired Student's *t* test (with Welch's correction as necessary). When one-way ANOVAs were performed, the Tukey post-test was used to determined significance of pairwise differences. $P \leq 0.05$ is denoted with \*, $\leq 0.01$ with \*\*, $\leq 0.001$ with \*\*\*, and $\leq 0.0001$ with \*\*\*\*; $P > 0.05$ is considered not significant.

**Supplemental Results**

Esterification of HEMA with LA, using EDC coupling, gave HEMA-LA monomer as a yellow oil (Supplementary Fig. 1a). HEMA-LA was characterized by $^1$H NMR spectroscopy, noting the vinyl protons at 5.4 and 6.1 ppm, and the characteristic signals from lipoic acid moiety in the alkyl region (~2.5 ppm) of the spectrum (Supplementary Fig. 1b, c).

Thiol concentration was quantified using Ellman's test with a typical calibration curve between the absorbance reading and thiol concentration (Supplemental Fig. 2a). The reduction time from disulfides to thiols was optimized by varying the reaction time from 2-10 h and the thiol concentration was measured immediately after each run. Also, two groups of gels were prepared in which one group was the swollen gels sitting in swelling solution for 5 days and the other was the swollen gels that underwent vacuum drying step for 48 h at room temperature. The result showed that the swollen gels had slightly higher free thiol production compared to the vacuum dried gels. It was determined that the optimal reduction time is 6 h by simply using swollen gels (Supplemental Fig. 2b). The reduction yield was also determined for each group of gels in all 3 sets (Supplemental Fig. 2c). The swelling behavior of reduced gels was also studied as defined in Equation (1). Supplemental Fig. 2d-f showed that reduced gels from all 3 sets exhibited deswelling over time post-reduction. The deswelling occurred after about 1 day after the reduction for Sets I and II (both composed of short co-monomer M1) and about 2 days for Set III (composed of long M1). This deswelling behavior indicated that thiol crosslinking occurred over time after the reduction. Moreover, the increase in gel stiffness as shown in Fig. 4 proved that this additional crosslinking was from inter-strand disulfide formation instead of cyclic disulfide reformation.